\documentclass{article}

\usepackage{arxiv}

\usepackage[backend=bibtex, sorting=none]{biblatex}
\bibliography{references} 

\usepackage[utf8]{inputenc} % allow utf-8 input
\usepackage[T1]{fontenc}    % use 8-bit T1 fonts
\usepackage{hyperref}       % hyperlinks
\usepackage{url}      
\usepackage{mathtools}
\usepackage{booktabs}       % professional-quality tables
\usepackage{amsfonts}       % blackboard math symbols
\usepackage{nicefrac}       % compact symbols for 1/2, etc.
\usepackage{microtype}      % microtypography
\usepackage{lipsum}
\usepackage{graphicx}
\usepackage{subcaption}
\graphicspath{ {./images/} }

\newcommand{\pt}{$p_{\mathrm T}$}
\newcommand{\ptj}{$p_{\mathrm T}^{j1}$}
\newcommand{\pty}{$p_{\mathrm T}^{\gamma1}$}
\newcommand{\Ht}{$H_{\mathrm T}$}
\newcommand{\MHt}{$H_{\mathrm T}^\mathrm{miss}$}
\newcommand{\decade}{DECADE}%{\textsc{DecADe}}
\newcommand{\myy}{$m_{\gamma\gamma}$}
\newcommand{\ptyy}{$p_\text{T}^{\gamma\gamma}$}
\newcommand{\yyjets}{$\gamma\gamma (+jets)$}

\title{\decade: Decorrelated anomaly detection triggers 
to enhance the low-mass discovery potential of the LHC}

\author{
 Noah Clarke Hall\\
  Centre for Data Intensive Science and Industry\\
  and Department of Physics and Astronomy\\
  University College London\\
  \texttt{noah.clarke-hall.22@ucl.ac.uk} \\
   \And
 Nikolaos Konstantinidis \\
  Centre for Data Intensive Science and Industry\\
  and Department of Physics and Astronomy\\
  University College London\\
  \texttt{n.konstantinidis@ucl.ac.uk} \\
}

\begin{document}
\maketitle
\begin{abstract}
At the ATLAS and CMS experiments at CERN's Large Hadron Collider, the rate of proton-proton collisions far exceeds the rate at which data can be recorded. 
A real-time event selection process, or "trigger", is needed to ensure that the data recorded contains the highest possible discovery potential. 
In the absence of hoped-for anomalies that would lead to the discovery of new physics, there is increasing motivation to develop dedicated, model-agnostic, anomaly detection triggers. 
A common approach is to use unsupervised machine learning (ML) to predict an event-by-event anomaly score, based on the momenta and multiplicity of reconstructed objects. 
Such anomaly scores often exhibit high correlation with existing trigger observables and thus exhibit a selection bias towards high-momentum anomalies. 
In this article, we introduce DECorrelated Anomaly DEtection (\decade), in which quantile regression is applied to the output of a pre-trained anomaly detection algorithm, guaranteeing the independence of the threshold on the anomaly score with respect to primary trigger observables. 
Thus, \decade\ provides efficiency in low-momentum regions of phase space not captured by existing triggers, boosting the trigger efficiency for low-mass phenomena that are inaccessible via primary triggers and current anomaly detection triggers. 
Quantile regression is implemented using decision tree ensembles, making \decade\ highly computationally efficient and therefore optimal for use both in software-based trigger systems and in FPGA-based hardware triggers. 
In both cases, we demonstrate that \decade\ would add an insignificant additional latency and resource cost to the hardware anomaly detection triggers currently in operation at ATLAS and CMS, as well as to those proposed for the High-Luminosity era of the Large Hadron Collider.
\end{abstract}

% keywords can be removed

\section{Introduction}
The Standard Model (SM) of particle physics has proved to be a remarkably successful description of fundamental particles and their interactions. The set of particles predicted by the SM was completed in 2012 with the discovery of the Higgs boson. However, the SM is not able to explain some major experimental observations in nature, such as astronomical evidence for dark matter (DM), the matter-antimatter asymmetry in the universe, or answer theoretical discrepancies, like the apparent fine-tuning of the mass of the Higgs boson. This motivates the continuing search for undiscovered, beyond-the-SM (BSM) particles and phenomena.

A large theoretical effort continues towards extensions to the SM that would include these phenomena. Some, such as supersymmetry (SUSY), provide a mathematically elegant framework, while predicting new particles with mass ranges and couplings to SM particles accessible to the Large Hadron Collider (LHC) at CERN. This spurred a range of dedicated model-dependent searches in theoretically motivated signatures, especially at mass scales of $\mathcal{O}$(TeV); however, no significant excesses have been observed at the time of writing. This absence of hoped-for anomalies has increased interest in model-independent approaches to event selection and analysis~\cite{ATLAS:2019gensearch, ATLAS:2023HggMI, ATLAS:2024unsup, CMS:2025dijetAD}, based on the compatibility of a given event with the background-only hypothesis. Such anomaly-detection (AD) techniques compromise performance for specific signal models in favour of increased efficiency to a range of model non-specific ``anomalies'', and are designed to complement targeted strategies. Unsupervised machine learning (ML) provides an ideal framework for constructing such anomaly detectors, capable of detecting anomalies effectively in the high-dimensional feature spaces common in LHC datasets. Such unsupervised methods are trained on a background-only hypothesis, making them independent of any choice of signal model.

The rate of proton-proton collisions at ATLAS and CMS greatly exceeds the rate at which data can be recorded, with only a small fraction of events ever being stored for use in analysis. As a result, events must be selected in real-time to ensure that the data recorded contains the highest possible discovery potential. The ATLAS and CMS trigger systems consist of two stages, an initial FPGA-based "hardware" trigger and a subsequent CPU-based "software" trigger. Individual event selection criteria are called "triggers", and typically consist of thresholds on physical observables, such as the multiplicity $N$ and transverse momentum \pt\ of reconstructed particles. The output data rate of the trigger system has a strict upper limit, and so triggers must be carefully tuned and combined to maximise efficiency for known signals and well-motivated anomalies. 

A frequent assumption in the design of triggers at ATLAS and CMS is that interesting collisions (signal events) produce particles that have on average higher mass or momenta than backgrounds, which dominate at lower mass- and momenta-scales. This bias is reflected in the design of the highest-rate triggers at ATLAS and CMS, which only select events above a given threshold in the \pt-$N$ space. Efforts to improve the efficiency of the triggers often focus on reducing these thresholds rather than incorporating other features such as the angular distributions of the particles in the event. However, there is renewed interest in BSM models that predict the existence of new physics at momentum scales below those currently accessible by conventional trigger strategies. Approaches such as trigger-level analysis and data-scouting aim to access these scales by bypassing the software trigger and record events at a higher frequency~\cite{CMS:2016scout, ATL-DAQ-PUB-2017-003, ATLAS:2018tla, CMS:2024scout}; however, the limited data rate requires that such events be noisier and lower-resolution than those that pass the entire trigger, partially offsetting any gain in sensitivity.

The efficiency of ML anomaly-detectors is likewise often restricted to high-momentum regions of phase space. This is a corollary of how the models are trained; the background rate falls sharply with \pt\ and $N$, so naturally events with high \pt\ and $N$ are more likely to be selected as anomalous. This pattern can be observed in the events selected by GELATO and AXOL1TL~\cite{gelato, axol1tl, axol1tl_resource}, the AD triggers currently in operation in the ATLAS and CMS hardware triggers respectively, with efficiency generally increasing with increasing mass-scale. As a result, these triggers exhibit a selection bias towards high-momentum events, with poor efficiency for softer anomalies. 

We address this by applying a correction to the output of a pre-trained baseline anomaly detector such that the conditional rate is independent of the primary trigger observables. This correction term is derived under a background-only hypothesis, preserving the signal model-independence of the AD trigger. This technique, which we term \decade, extends trigger efficiency to low-mass and low-momentum regions of phase space. We demonstrate the potential physics benefit from \decade\ by studying the trigger efficiency for low-mass photon pairs, finding that efficiency improves by a large factor in regions of phase space where primary triggers are insensitive. In addition, we show that \decade\ could be deployed in both hardware and software triggers, without adding latency or significant computational cost to GELATO and AXOL1TL. While this work focuses on jets and photons as a case study, we emphasise that \decade\ is fully agnostic to the reconstructed objects included in the event and can be applied to a complete trigger menu containing leptons or missing energy. 

\section{Simulated datasets}
%Basic description of ATLAS / CMS
This study uses Monte Carlo simulations of proton-proton collisions sourced from the public ATLAS Open Data dataset~\cite{ATLAS2024DAOD}. Events are simulated at $\sqrt{s}=13$~TeV, and correspond to data-taking conditions during 2015 and 2016. A total bunch-crossing rate of 40 MHz is assumed throughout. To demonstrate the efficacy of \decade\, we implement a simplified trigger system where only $R=0.4$ jets and photons are reconstructed. Reconstructed jets (photons) are required to have \pt\ $>20 (10)$ GeV, and jets within $\Delta R<0.4$ of a photon are removed. The 6 highest-\pt\ jets and 4 highest-\pt\ photons are selected in each event, padded with zeros to account for missing particles. The \pt\ of all reconstructed particles are scalar-summed (\Ht) and vector-summed (\MHt), with jet multiplicity $N_\text{j}$ and photon multiplicity $N_\gamma$ also taken. Particle momenta are represented in cylindrical coordinates $(p_\text{T}, \eta, \Delta\phi)$, where $\Delta\phi$ is the azimuthal angle between the particle and the \MHt\ vectors. We perform an anomaly search in an "enhanced-bias" region defined by $N_\gamma\geq1$.

% Simulated events are generated at $\sqrt{s}=13$~TeV using \textsc{MadGraph5+Pythia8}, with the detector response simulated using the default ATLAS card in \textsc{Delphes}. A minimum-bias dataset is produced by generating 1M events for the inclusive SM process $pp\rightarrow jj$ at leading-order, where $j\in\{g, u, d, s, c, b\}$. In order to generate a set of anomalies with diverse mass-scales, we generate the BSM processes $pp\rightarrow Z'\rightarrow jj$ and $pp\rightarrow Z' Z' \rightarrow jj\nu\nu$, where $Z'$ is a vector boson implemented using the \textsc{VPrimeNLO} model~\cite{Fuks_2017}. The mass of the $Z'$ is drawn from $M_{Z'}\in\{15, 20, 30,60, 100, 300, 1000\}$ GeV, with 50,000 events generated for each mass point and process. Jets are reconstructed using the anti-$k_t$ algorithm in \textsc{FastJet}, with $R=0.4$ and \pt\ $>10$ GeV.

We construct a "minimum-bias" dataset which represents the overwhelming majority of events accepted by jet and photon triggers, consisting of 10M $pp\rightarrow jets$ events and 1M $pp\rightarrow \gamma (+jets)$ events. Trigger efficiency is evaluated using SM $pp\rightarrow\gamma\gamma (+jets)$, a non-resonant process covering a broad spectrum of event topologies. Events for $pp\rightarrow\gamma\gamma (+jets)$ are required to have $N_\gamma\geq2$. This sample allows trigger efficiency to be profiled over the invariant mass \myy\ and vector-summed transverse momentum \ptyy\ of the two highest-\pt\ photons for a wide range of values. These observables are of particular relevance to BSM models that predict low-mass particles decaying to pairs of photons. Recent ATLAS and CMS analyses have searched for such resonances using Run 2 data~\cite{atlas_lowmass_diphoton}\cite{cms_lowmass_diphoton}. The sensitivity of these analyses falls with the invariant mass of the mediator and is limited to highly boosted topologies by photon trigger thresholds. Thus, additional efficiency at low \myy\ and \ptyy\ could greatly expand the sensitivity of such analyses. Without access to samples for these specific BSM models, we use $pp\rightarrow\gamma\gamma (+jets)$ to demonstrate the increased sensitivity that \decade\ offers to low-mass di-photon processes. 
%Detailed information on the datasets used is listed in Appendix A.
%Other resonant signals containing combinations of jets and photons are added, such as SM ggF and VBF $H\rightarrow\gamma\gamma$. 

The reconstructed offline objects used in this study are of higher purity and momentum-resolution than those input to real trigger algorithms. As a result, this study is not intended to accurately model the performance of the ATLAS or CMS trigger systems, serving only as proof-in-principle of the efficacy of \decade\ as a method. We implement a simplified jet- and photon-only primary trigger menu, to which an anomaly detection trigger is then added. The primary triggers implemented are leading-jet \ptj\ $>100$ GeV and leading-photon \pty\ $>30$ GeV, which are based on the J100 and EMVHI30 ATLAS L1 triggers in Run 2~\cite{phase1tdr}. An additional anomaly trigger is added to the menu that accepts the top $0.5\%$ of events that fail both primary triggers, equivalent to an estimated unique rate of 1.35 kHz. By comparison, the estimated rates for \ptj\ $>100$ GeV and \pty\ $>30$ GeV are estimated as 18.5 kHz and 3.28 kHz respectively.

\section{Decorrelated anomaly triggers}
We describe an arbitrary trigger menu on some primary observables $x=\{x_i\}$, to which we wish to add an ML discriminant $y$. We let $b(x, y)$ be the density of background events, and define the marginal density $b_X=\int b(x,y) dy$. For a single trigger observable $x_i$, a threshold $X_i$ can be chosen such that a given fraction of background events have $x_i>X_i$ and are accepted. In the case of multiple parallel triggers, an event is accepted if it passes any trigger threshold. In this scenario, the trigger rate is given by
\begin{equation}
    R_X = R_0\int_{V_X} b_X(x) dx
\end{equation}
where 
\begin{equation}
    V_X=\bigcup_i \{x_i>X_i\}
\end{equation}
is the volume of phase space accepted by the trigger and $R_0$ is the total rate of collisions. Since $b(x, y)\geq0$ for all $(x,y)$, both rate $R_X$ and accepted phase space $V_X$ increase monotonically with falling $X_i$. The combined rate from all multiple parallel triggers is always less than or equal to the sum of the rates from each component trigger, since an event can be accepted by multiple triggers. When added to the ensemble, the new trigger $y$ adds a unique rate
\begin{equation}
    \Delta R_Y = R_{XY} - R_X \leq R_Y.
\end{equation}
This unique rate can be set by fixing the thresholds $\{X_i\}$ and scanning over the threshold $Y$ until the unique rate reaches the allocated budget.
% Due to the monotonicity of $R_X$ and $V_X$, we must also have a unique phase space
% \begin{equation}
%     \Delta V_Y = V_{XY} - V_X \leq V_Y.
% \end{equation}
% The non-unique rate $R_Y - \Delta R_Y\geq 0$ therefore implies a corresponding lost unique phase space over $y$, $V_Y - \Delta V_Y\geq 0$. The greater the degree of positive correlation between $X$ and $Y$, the smaller the unique rate and therefore the unique phase space. 
% \begin{figure}
%     \centering
%     \includegraphics[width=0.9\linewidth]{figures/decade.drawio.pdf}
%     \caption{\decade\ applies a }
%     \label{fig:enter-label}
% \end{figure}
\subsection{Quantile regression}
 % However, if $x$ and $y$ are positively correlated, the unique phase space added will still be largely adjacent to that accepted by the primary triggers. Thus, anomalous events for which $y>Y-\delta Y$ are still likely to kinematically resemble those selected by the primary triggers, and so the profile of signal events that the trigger is sensitive to does not change. 
% The second disadvantage is that fixing $\{X_i\}$ and relaxing $Y$ requires that $Y$ take priority in accessing any extra available rate. With the sensitivity of numerous targeted LHC analyses depending on the thresholds set on specific primary triggers, priority for $y$ is difficult to justify.

\begin{figure}
    \centering
    \includegraphics[width=\linewidth]{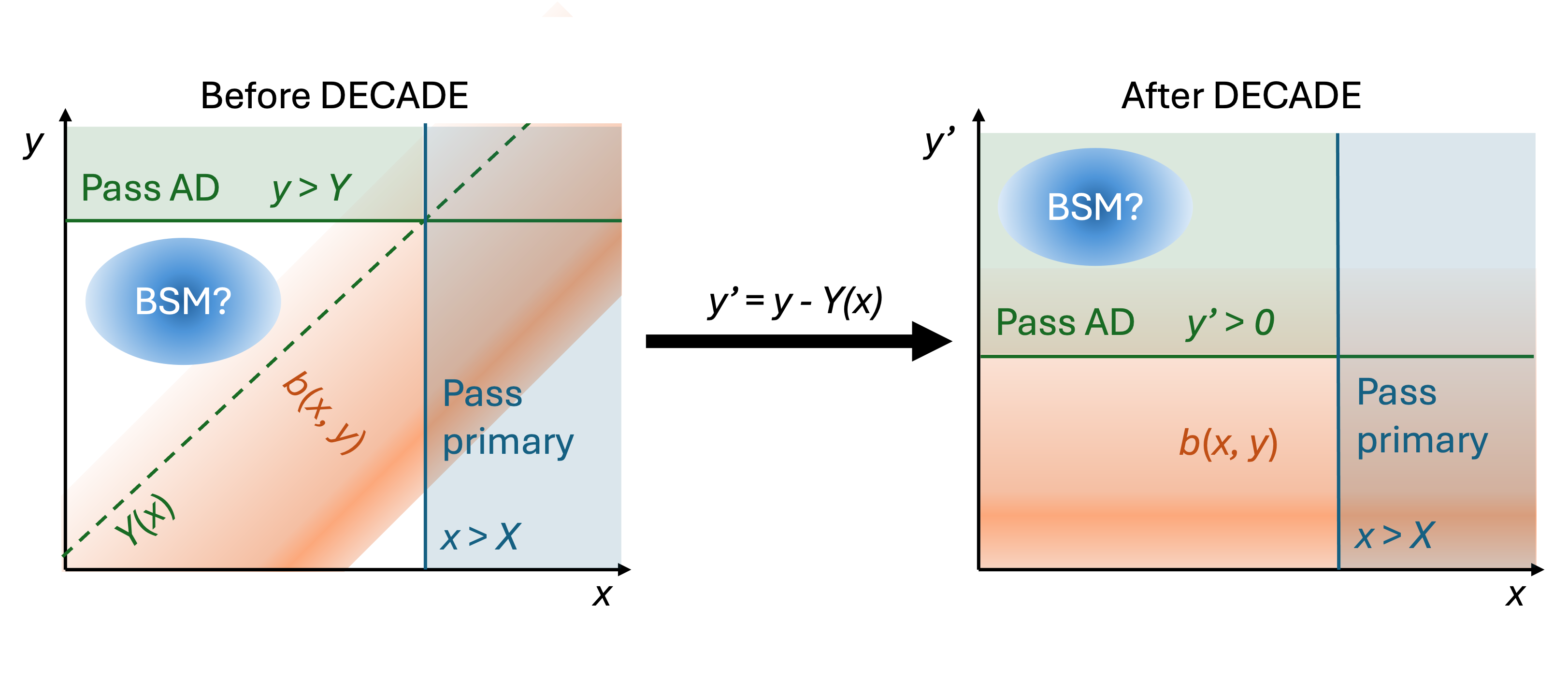}
    \caption{\decade\ guarantees a uniform conditional rate, improving efficiency for soft BSM signals.}
    \label{fig:decade_diagram}
\end{figure}

Our proposal is to apply a correction to the ML discriminant that breaks the correlation between $R_X$ and $R_Y$. We achieve this by transforming $y$ under the transformation $y\rightarrow y' = y - Y(x)$, where $Y(x)$ is the conditional quantile satisfying
\begin{equation}
    \int_{y>Y(x)} b(y | x)\, dy = q,
\end{equation}
where $q\in [0, 1]$ and is independent of $x$. Here, $q$ is the target conditional background efficiency, chosen to match the allowed rate budget for the anomaly trigger. Here we assume $b(y | x)$ is continuous in $y$ at $Y(x)$, such that $q$ is well-defined. Note that since $Y(x)$ is unique and a scalar, the transformation $y\rightarrow y' = y - Y(x)$ preserves the discrimination power of $y$ for a given $x$. The condition $y>Y(x)$ is equivalent to the condition $y'>0$ by construction, and since $dy=dy'$, we can write the conditional rate as
\begin{equation}
    R_{Y'=0|X} = R_0 \int_{y'>0} b(y' | x)\, dy' = R_0 \int_{y>Y(x)} b(y | x)\, dy = q R_0 .
\end{equation}
Thus the condition $y'>0$ is guaranteed to accept a uniform fraction $q$ of background events for any given choice of $x$. 
% If $R_X$ and $R_Y$ are positively correlated, as is common for an ML discriminant, then the transformation $y\rightarrow y'$ must guarantee a larger unique rate and thus a larger unique phase space. 
A diagram illustrating the potential effect of \decade\ for a single primary trigger is shown in Fig.~\ref{fig:decade_diagram}.

With access only to samples drawn from $b(x,y)$, we can approximate $Y(x)$ by constructing an estimator that minimises the pinball loss~\cite{Steinwart_2011}. While the choice of estimator for $Y(x)$ is largely free, we select a boosted decision tree (BDT) due to their excellent performance on tabular data and high computational efficiency. The BDTs are implemented using the \textsc{XGBoost} package~\cite{Chen_2016} using a learning rate of 0.3, a maximum tree depth of 4 and a maximum ensemble size of 100. Early stopping is applied to prevent overfitting and to minimise the ensemble size.

\section{Results}
\subsection{Baseline model}
We train an unsupervised autoencoder neural network as a baseline anomaly detector, consisting of some fully-connected layers implemented using the \textsc{TensorFlow} package. The fully-connected layers contain $\{32, 16, 8, 3, 8, 16, 32\}$ neurons respectively, with the ReLU activation function applied between layers. The autoencoder is trained to compress enhanced-bias events to a 3-dimensional latent space and then reconstruct it such that the mean-squared error (MSE) between input and output is minimised. Outliers tend to be poorly reconstructed, and so the log of event-wise MSE can be taken as an AD discriminant $y$. The MSE is masked to exclude missing features, meaning that missing particles do not contribute to the loss or discriminant. The autoencoder is trained on the features $\{N_\text{j}, N_\gamma, H_\text{T}, H_\text{T}^{miss}, p_\text{T}^i, \eta^i, \Delta\phi^i \}$, where $i$ denotes the 6 highest-\pt\ jets and 4 highest-\pt\ photons in the event, ordered first by particle type and then by \pt. Each feature is independently normalised by subtracting the mean and dividing by the standard deviation, ignoring missing particle momenta which would otherwise bias the variance. The model is trained for 200 epochs using the Adam optimiser with a learning rate of $10^{-3}$. A 50-50 train-validation split is used, with a test set being drawn from within the training set. The same splitting is used to train \decade, ensuring that the validation set remains unseen during training.

% The second is a supervised BDT classifier trained to separate enhanced-bias events from VBF $H\rightarrow\gamma\gamma$ production. This signal process has a highly distinct event topology, and this model is an example . The BDTs are implemented in \textsc{XGBoost}, using the default hyperparameters. Such a classifier serves to demonstrate that \decade\ can be applied to both unsupervised and supervised models, able to convert even the most signal-specific discriminant into a robust and performant anomaly-detector. Supervised models typically out-perform their unsupervised counterparts

\subsection{Trigger rates}
\begin{figure}
    \centering
    \begin{subfigure}{\textwidth}
    \centering
    \includegraphics[width=\linewidth]{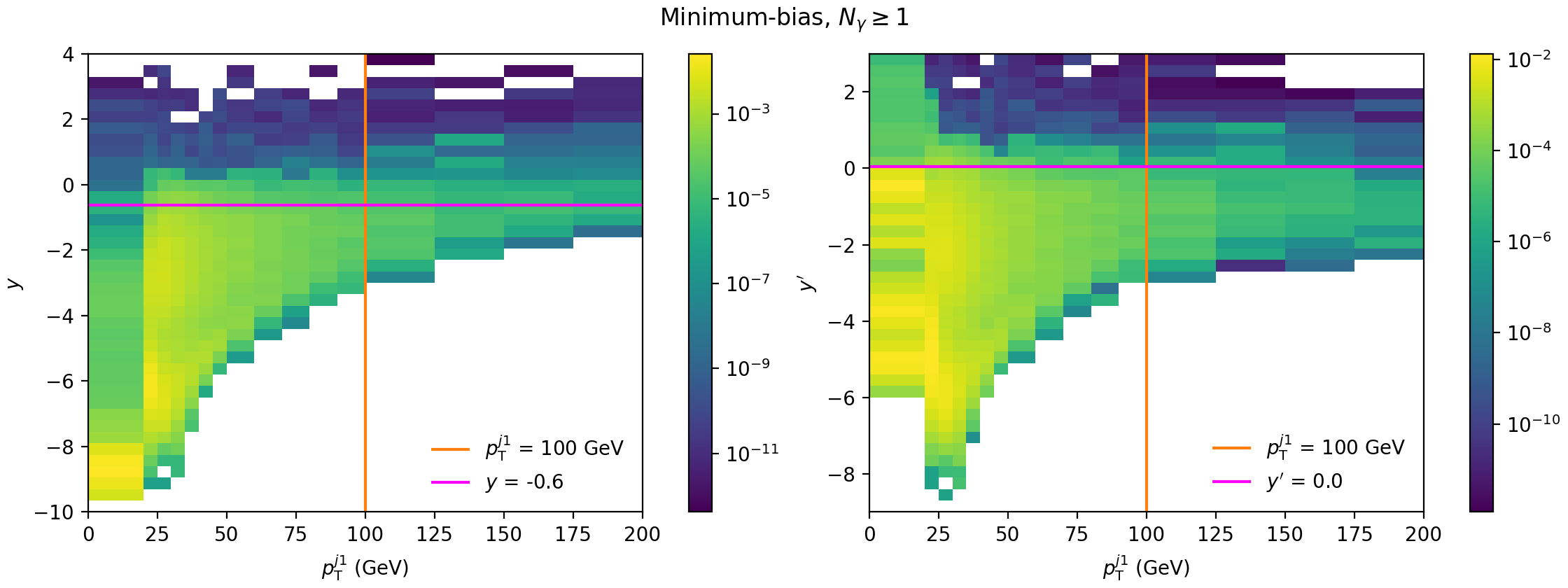}
    \end{subfigure}
    \begin{subfigure}{\textwidth}
    \centering
    \includegraphics[width=\linewidth]{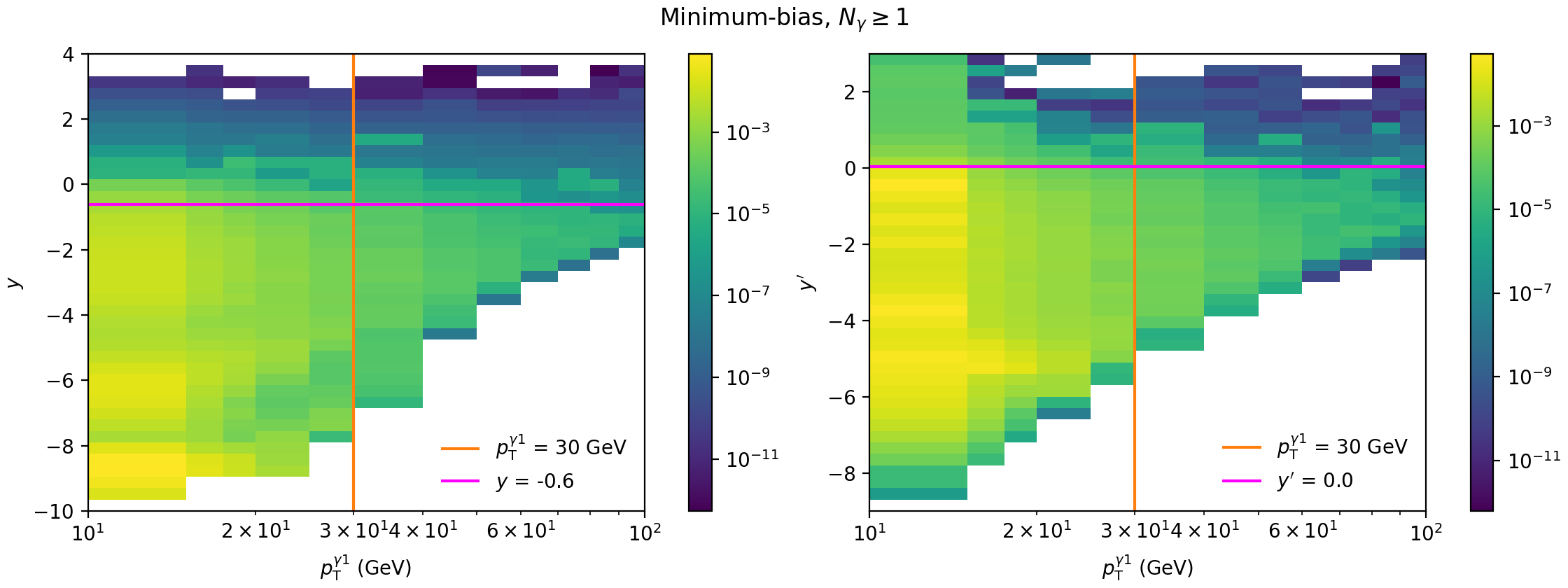}
    \end{subfigure}
    \caption{Distributions of minimum-bias events as a function of \ptj\ and \pty\ before and after \decade\ is applied.}
    \label{fig:minbias_2d}
\end{figure}
Figure~\ref{fig:minbias_2d} shows the distribution of validation minimum-bias events as a function of the primary trigger observables \ptj\ and \pty\ for the AD discriminant with and without \decade. The corresponding thresholds are also shown. The fraction of events passing the baseline AD trigger increases with \ptj\ and \pty. The effect of \decade\ is clearly visible, with the distortion introduced by $y\rightarrow y'=y-Y(x)$ producing a more uniform conditional rate. Events with lower \ptj\ and \pty\ tend to have a more positive shift, indicating that \decade\ is allocating additional rate to softer event topologies. The total rates for the autoencoder without and with \decade\ are 1.62 kHz and 1.42 kHz respectively. With the unique rate fixed at 1.35 kHz, the total rate with \decade\ is 95\% unique compared to 83\% without, demonstrating how \decade\ reduces the overlap between AD and primary triggers. 

\subsection{Low invariant-mass photon pairs}
\begin{figure}
    \centering
    \begin{subfigure}{0.49\textwidth}
    \centering
    \includegraphics[width=\linewidth]{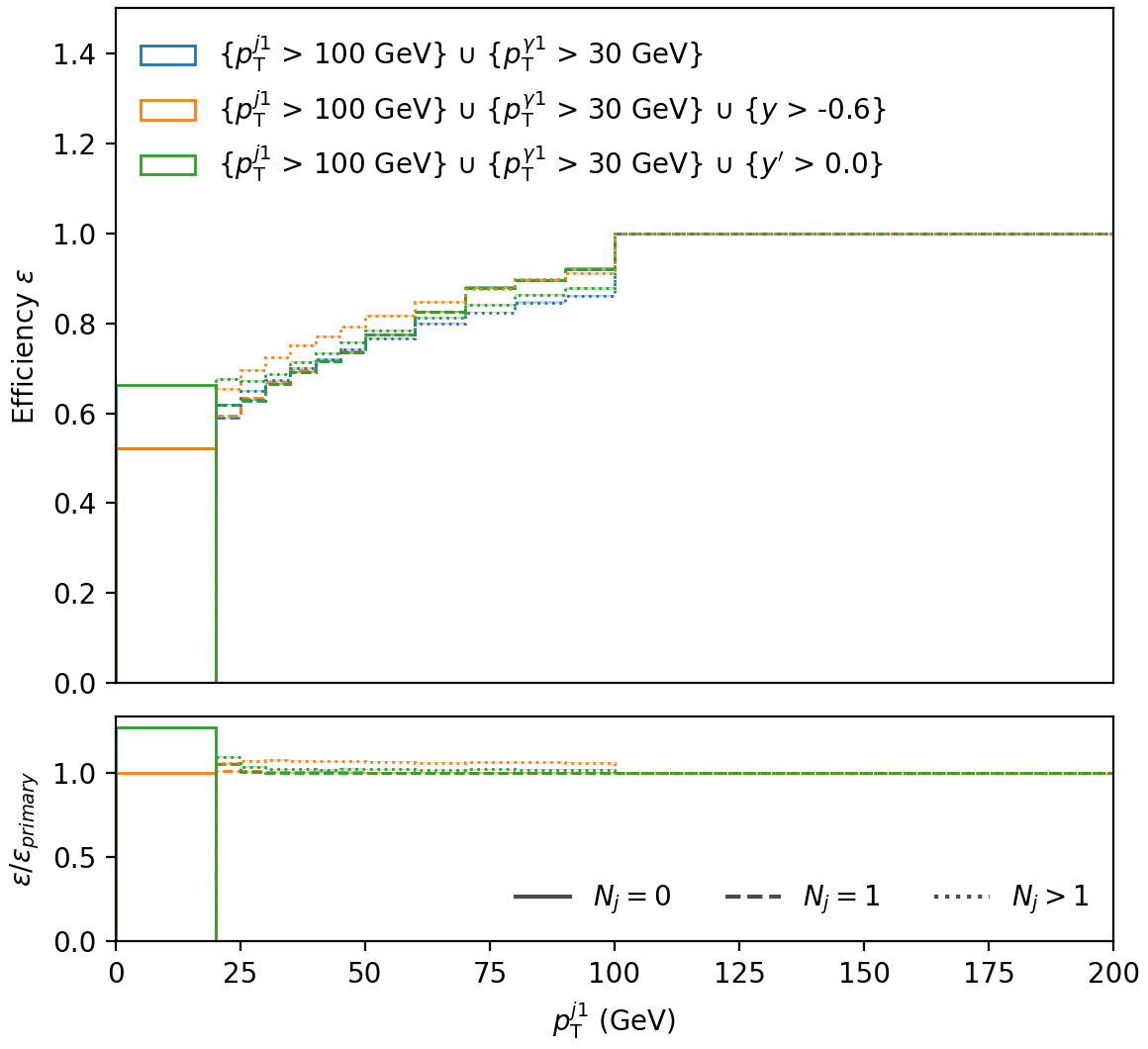}
    \end{subfigure}
    \begin{subfigure}{0.49\textwidth}
    \centering
    \includegraphics[width=\linewidth]{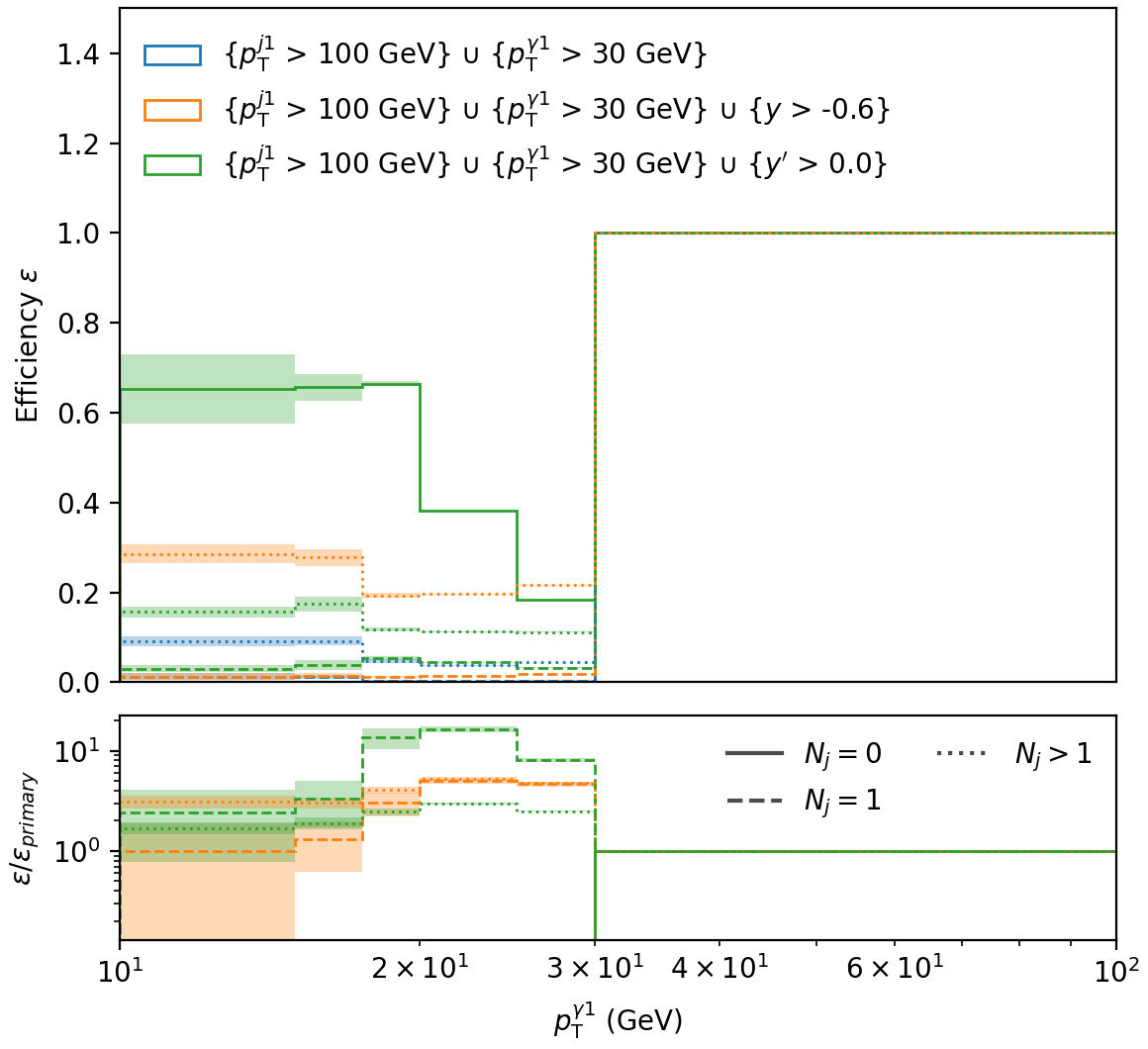}
    \end{subfigure}
    \begin{subfigure}{0.49\textwidth}
    \centering
    \includegraphics[width=\linewidth]{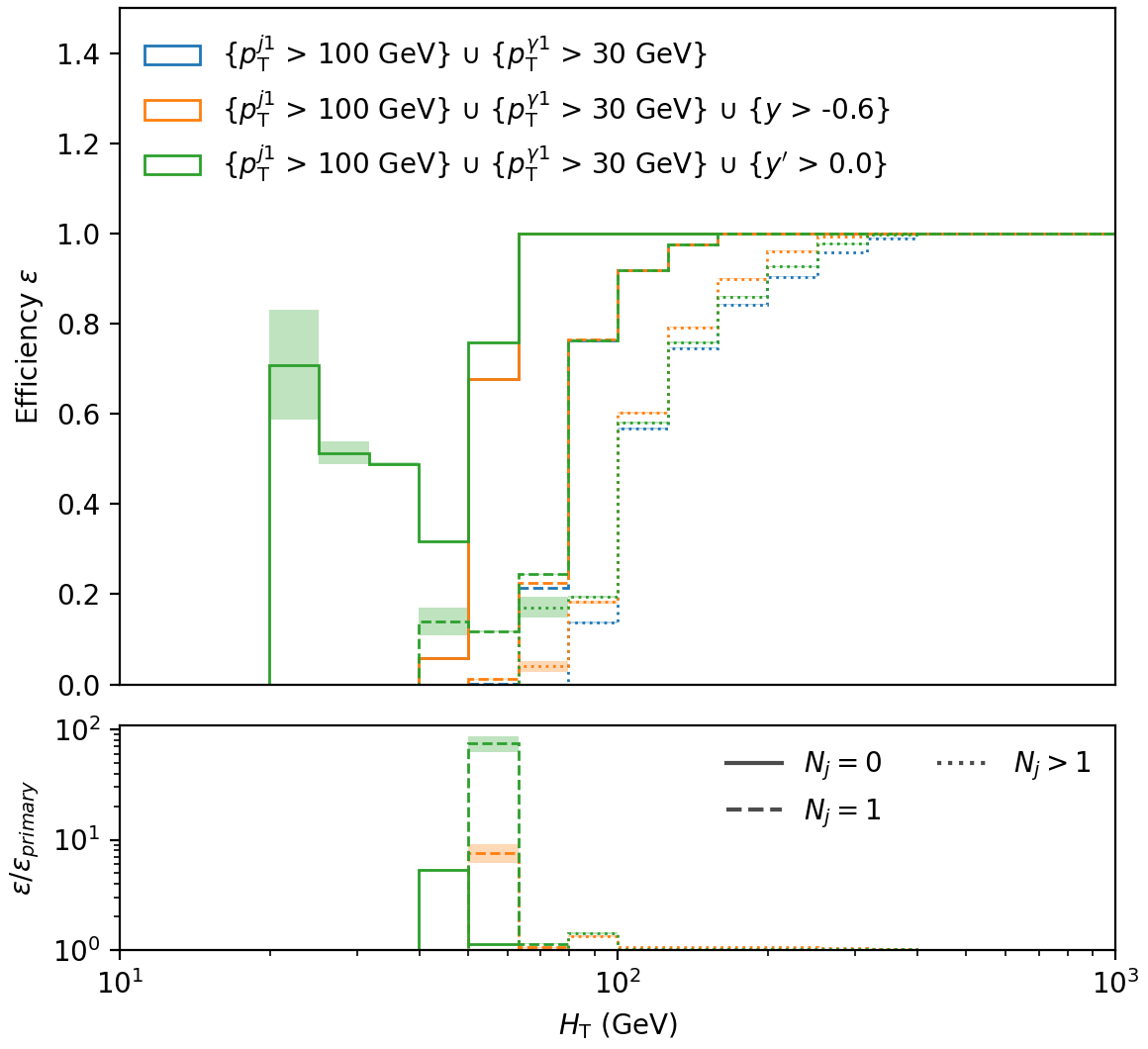}
    \end{subfigure}
    \begin{subfigure}{0.49\textwidth}
    \centering
    \includegraphics[width=\linewidth]{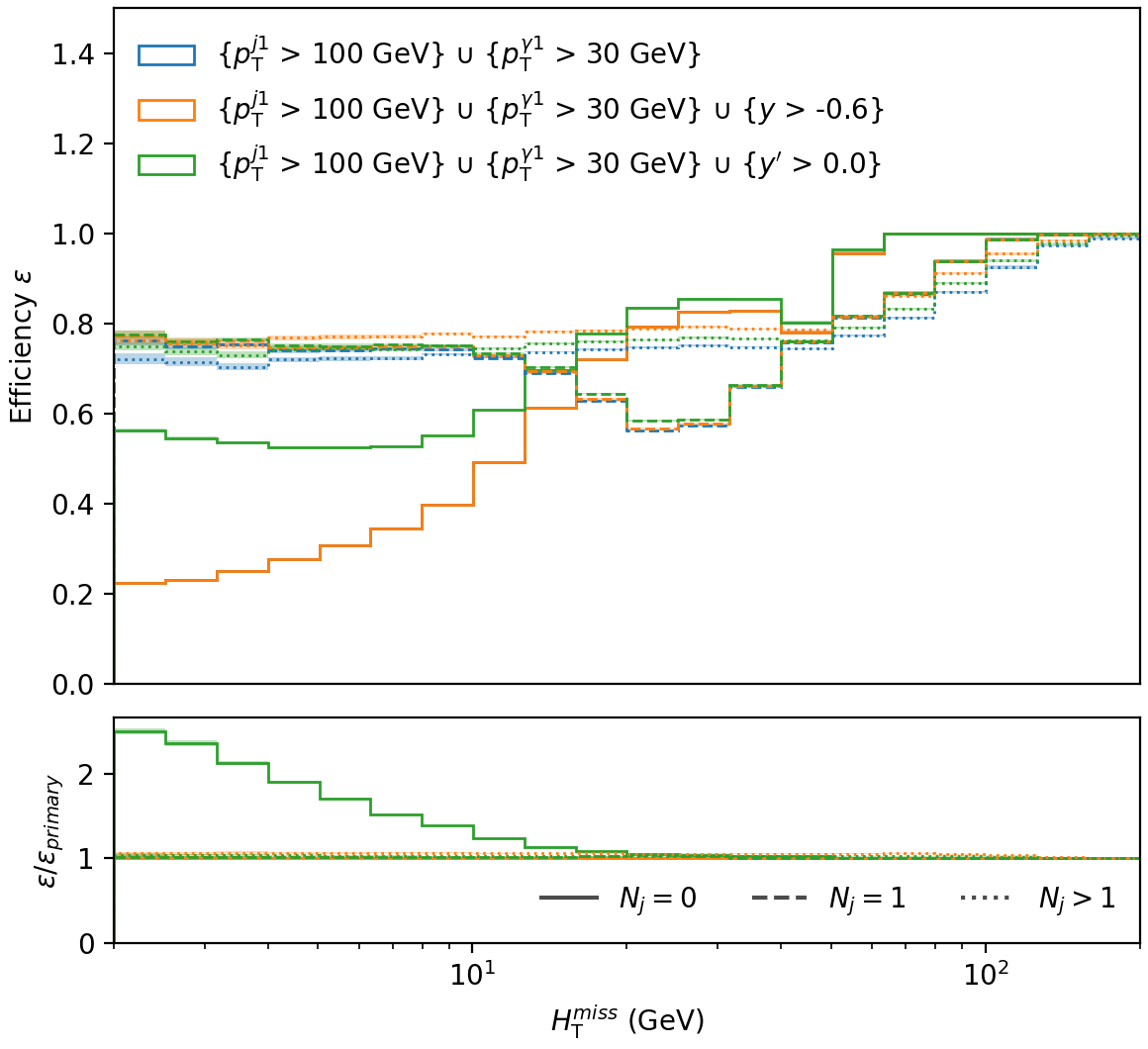}
    \end{subfigure}
    \begin{subfigure}{0.49\textwidth}
    \centering
    \end{subfigure}
    \begin{subfigure}{0.49\textwidth}
    \centering
    \includegraphics[width=\linewidth]{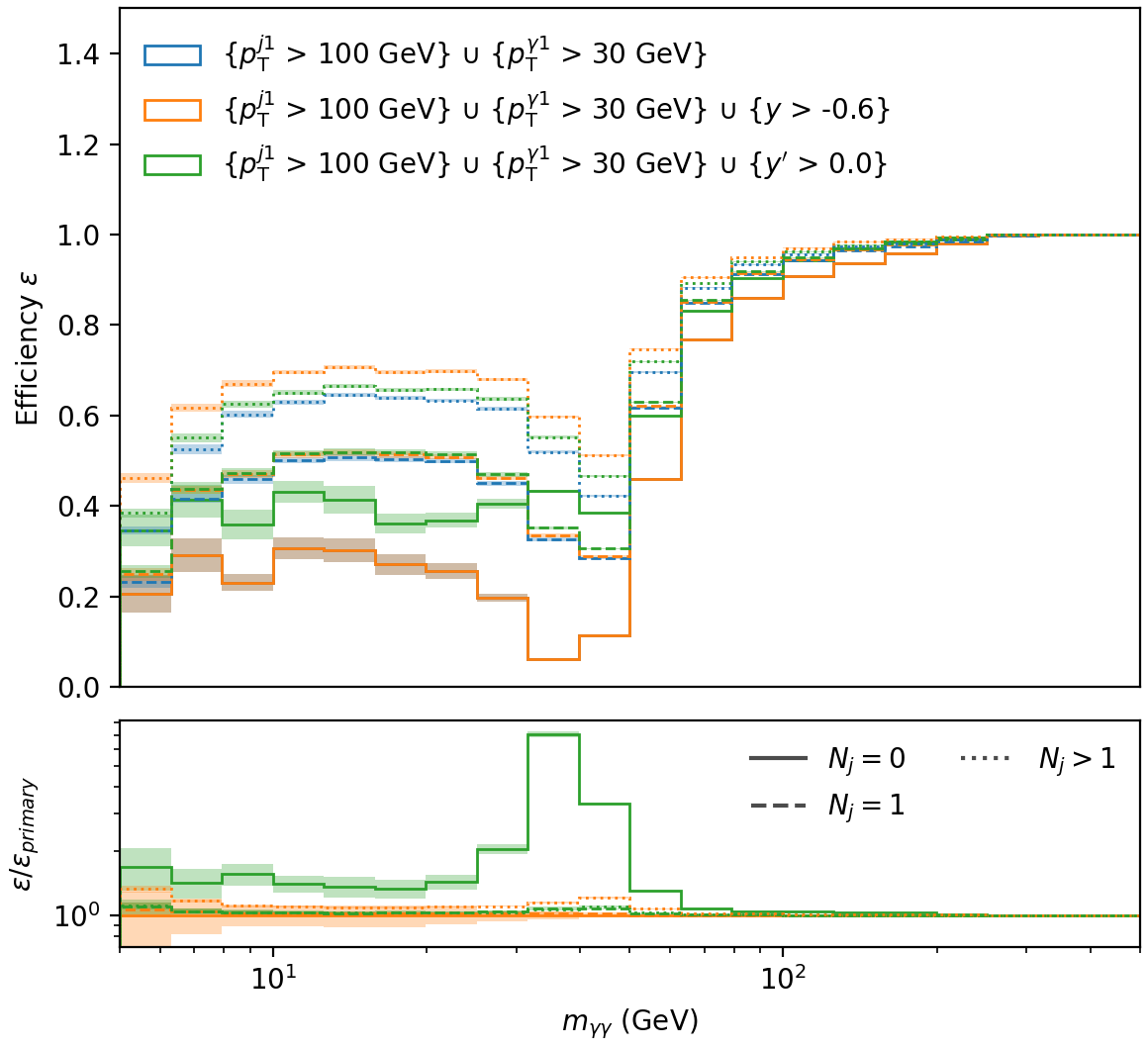}
    \end{subfigure}
    \begin{subfigure}{0.49\textwidth}
    \centering
    \includegraphics[width=\linewidth]{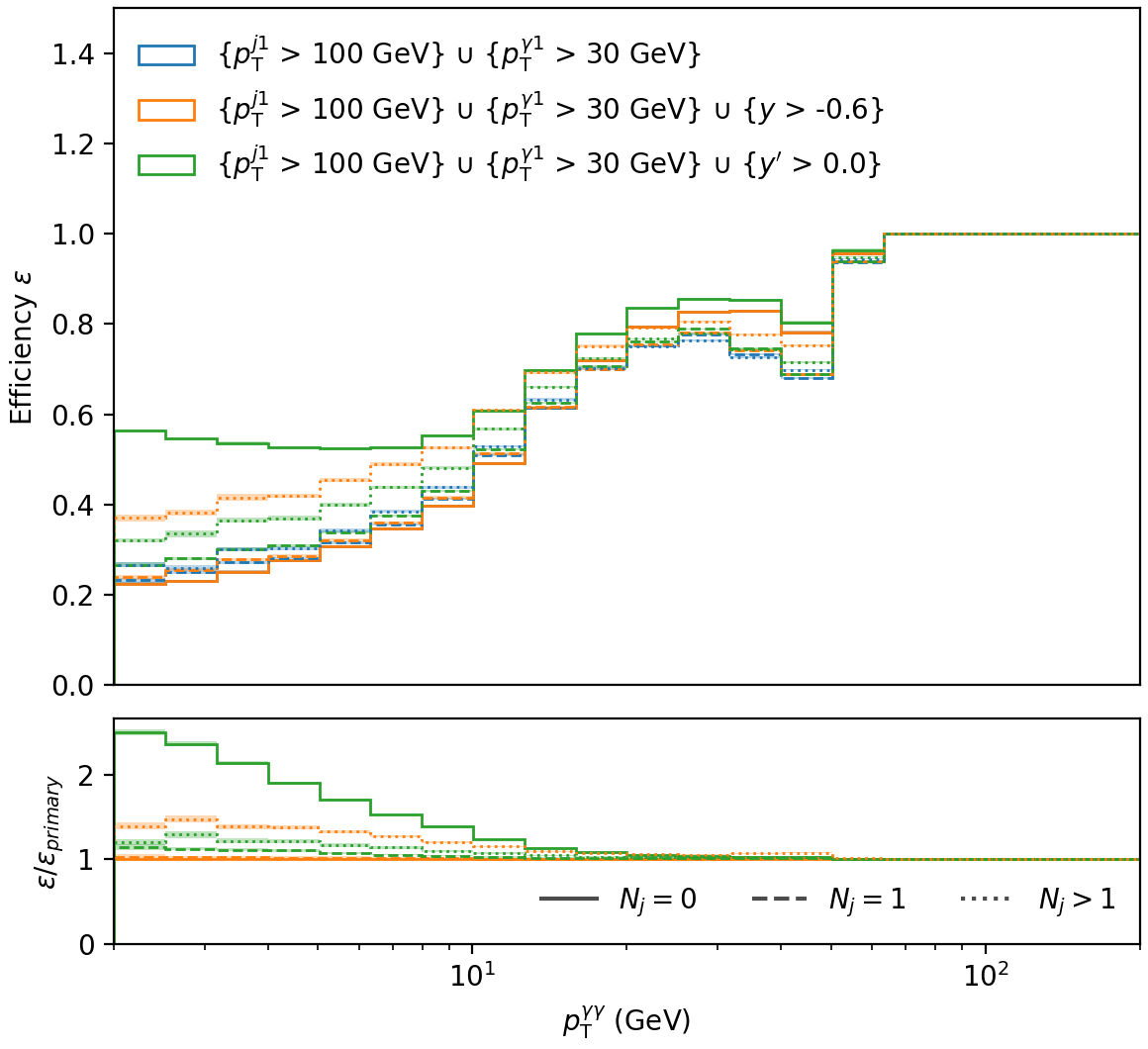}
    \end{subfigure}
    \caption{Combined trigger efficiency of \yyjets\ as a function of various explanatory observables.}
    \label{fig:turn_ons}
\end{figure}

The additional efficiency to low invariant-mass photon pairs achieved with \decade\ can be characterised by marginalising \yyjets\ efficiency over a set of explanatory observables. For each AD trigger, we measure the combined efficiency of the AD and primary triggers and compare it to that obtained from only the primary triggers. This allows us to compare the effect of overlap between triggers and provide a complete profile of the sensitivity of the trigger as a whole. The explanatory observables chosen include the primary observables \ptj and \pty, to which we add \Ht, \MHt, \myy\ and \ptyy. Efficiency is categorised by $N_j$ to show how \decade\ performs on different event topologies. The $N_j=0$ category contains events for which only the photon pair is reconstructed. The category $N_j=1$ contains events in which the photon pair may be boosted by a recoiling jet. All other events are combined into the $N_j>1$ category. The marginal \yyjets\ efficiency of the combined triggers are shown for each explanatory observable in Fig.~\ref{fig:turn_ons}. We evaluate the statistical uncertainty from finite statistics, indicated by shaded bands. While not explicitly included as inputs to the BDT, the effect of \decade\ extends to all observables by way of correlation with \ptj\ and \pty. \decade\ greatly increases efficiency for $N_j\leq1$, with this improvement being most pronounced for $N_j=0$. For $N_j>1$, \decade\ increases efficiency where \Ht\ $<100$ GeV, and while slightly reduced where \Ht\ $>100$ GeV, \decade\ still improves upon the primary-only trigger scenario. The relative gain in efficiency from \decade\ can be extreme in some regions of phase space, producing high efficiency where the primary trigger has low or zero efficiency. By contrast, efficiency gains from the baseline model largely follow the pattern of the primary triggers. 

%\subsection{Other signals}
% Show range of other signals (ggF Hyy, VBF Hyy, HZ-->yyqq), showing that efficiency does not decrease significatnly

\subsection{Deploying DECADE}
\decade\ can be deployed in both hardware and software triggers without adding significant computational resource and latency. As the correction term $Y(x)$ is independent of the prediction of the baseline model $y$, both can be calculated in parallel within the same latency window. This is particularly useful in the hardware trigger, due to the more stringent latency constraints. The BDT is relatively compact, with an ensemble size of 22 and 316 total split nodes. We benchmark the inference time of the trained BDT quantile regressor for a single event on a CPU and an FPGA. The CPU implementation is benchmarked on an Intel Xeon Gold 6148 for a single thread with a 2.4 GHz clock. The mean latency is measured to be $150\pm15$ $\mu$s, which is insignificant compared to the $\mathcal{O}(100\text{ ms})$ total latency of the software trigger. Latency could be reduced further by using multiple threads or batched inference. 

The FPGA implementation is benchmarked on a Xilinx Ultrascale+ XCVU9P with 250 MHz clock. The conversion of the BDT to firmware is performed using the \textsc{Conifer} package~\cite{Summers_2020}. The firmware uses a 12-bit and 16-bit fixed-point representation for the node thresholds and leaf weights respectively. The reduced precision of the leaf weights produces a slight numerical disagreement between the FGPA and CPU implementation, but this has no measurable effect on physics performance. Resource usage and latency is estimated by performing C-synthesis in \textsc{VitisHLS}. The latency is estimated to be just 8 ns, compared to the latency budgets for AXOL1TL and GELATO of 50 ns and 25 ns respectively~\cite{axol1tl_resource,gelato}. Resource usage is estimated to be $10.5$k look-up tables (LUTs) and 630 registers, $<1\%$ of board resources. Note that this is a conservative upper limit, as previous studies have reported that C-synthesis overestimates LUT usage by a large factor when compared to more realistic simulations~\cite{Summers_2020}. We expect resource usage of the final BDT firmware to be insignificant compared to the $\mathcal{O}(30\text{k})$ LUTs allocated AXOL1TL and GELATO~\cite{axol1tl_resource, gelato}.

\section{Discussion}
While this work focuses on low-mass photon pairs as an illustrative example of the physics impact of \decade\, we emphasise that \decade\ is fully agnostic to the objects included. We foresee \decade\ being used within the context of a full trigger menu that includes leptons and missing energy, providing far-ranging improvements in sensitivity for analyses whose search space falls outside of that accessible by both primary triggers and conventional AD. 
% Since unique rate decreases as more triggers are added to the menu, we expect the proportional gain from \decade\ to be even greater in a realistic full menu than in our simplified study. 
This makes it a promising addition not only for special-case triggers but as a systematic enhancement to the overall trigger menu design. 

We also emphasise that \decade\ is fully agnostic to the choice of base discriminant, so long as $b(y | x)$ is continuous in $y$ at $Y(x)$, such that the conditional quantile $q$ is well-defined. This property is satisfied not just by dense autoencoders, but also by variational autoencoders such as GELATO and AXOL1TL. Pseudo-continuous discriminants such as decision tree ensembles may also satisfy this, if the number of parameters is large enough. Crucially, the physics benefits are not predicated on \decade\ fully converging on the true conditional quantile, as any increase in the uniformity of the conditional rate will add trigger efficiency for soft anomalies. For sparse or high-dimensional $x$, additional regularisation may be useful to guarantee that the BDT can well-estimate the conditional quantile. Here some dimensionality reduction may help, for example via principal component analysis. The BDT could also be switched for a neural network, as these are known to be more robust on sparse datasets~\cite{blanchard2021representationpowerneuralnetworks}.

\decade\ is designed to complement the baseline AD trigger, enabling incremental integration without retraining the existing discriminant. The lack of additional latency and small resource usage  greatly facilitates the deployment of \decade, especially within a hardware trigger where resource and timing constraints are most stringent. This enables \decade\ to improve trigger performance with minimal impact on current systems.

% The use of BDTs as quantile regressors works well for a simplified set of primary triggers, such that $x$ is low-dimensional. However, BDTs are susceptible to the "curse of dimensionality", especially when regressing extreme quantiles in regions where background is sparse. This challenge will need to be overcome if future work is to implement \decade\ with an expanded set of primary triggers. Dense neural networks are inherently less susceptible to the dimensionality problem~\cite{blanchard2021representationpowerneuralnetworks}, and may guarantee a smoother $Y(x)$ than tree-based approaches. Although less computationally efficient, the use of neural networks does not disqualify \decade\ from being implemented in either the hardware or software triggers, as both AXOL1TL and GELATO are examples of neural networks that have been successfully implemented in a hardware trigger. The problem can be further mitigated by omitting the primary triggers which are weakly correlated with the base AD trigger.

\section{Summary}
At the LHC’s ATLAS and CMS experiments, the enormous collision rate necessitates an efficient real-time trigger system to maximize discovery potential. With no guaranteed anomalies on the horizon, there is growing interest in model-agnostic, unsupervised anomaly detection triggers. 
However, existing approaches exhibit a selection bias towards high-momentum anomalies at the expense of efficiency for softer anomalies.
We resolve this by introducing DECorrelated Anomaly DEtection (\decade), which uses quantile regression on a pre-trained anomaly score to ensure trigger thresholds are independent of primary observables. This approach extends trigger efficiency into low-momentum phase space regions, enhancing sensitivity to low-mass phenomena. Implemented with decision tree ensembles, \decade\ is highly efficient, making it well-suited for both software- and FPGA-based triggers. Our studies show that \decade\ can be integrated with minimal latency and resource overhead, making it a practical and powerful addition to existing trigger systems at ATLAS and CMS, including future upgrades for the High-Luminosity LHC.

\section*{Acknowledgements}
We acknowledge the work of the ATLAS Collaboration to record or simulate, reconstruct, and distribute the Open Data used in this paper, and to develop and support the software with which it was analysed.

We gratefully acknowledge the support of the UK's Science and Technology Facilities Council (STFC). N.C. is supported by the STFC UCL Centre for Doctoral Training in Data Intensive Science (ST/W00674X/1) and by departmental and industry contributions.

This work used computing equipment funded by the Research Capital Investment Fund (RCIF) provided by UKRI, and partially funded by the UCL Cosmoparticle Initiative.
% \appendix
% \section{Datasets}
% \begin{table}[h]
%     \centering
%     \begin{tabular}{c|c|c|c}
%     Process &  DSID &  Events sampled & Notes \\
%     \hline
%     $jets$ & 364700 - 364704 & & Minimum-bias\\
%     $\gamma (+jets)$ & 423099 - 423104 & \\
%     \yyjets \\
%     \end{tabular}
%     \caption{Caption}
%     \label{tab:datasets}
% \end{table}

%\bibliographystyle{unsrtnat}

\printbibliography


\end{document}